\documentclass[conference]{IEEEtran}
\IEEEoverridecommandlockouts
\usepackage{cite}
\usepackage{float}
\usepackage{amsmath,amssymb,amsfonts}
\usepackage{graphicx, enumitem}
\usepackage{textcomp}
\usepackage[ruled,vlined,linesnumbered]{algorithm2e}
\usepackage[table,dvipsnames]{xcolor}
\usepackage{tikz}
\usepackage{bm}
\usepackage{makecell}
\usepackage{booktabs}
\usepackage{multirow}
\usepackage{url}
\usepackage{subfigure}
\usepackage{tikz}
\usepackage[none]{hyphenat}
\SetKwIF{If}{ElseIf}{Else}{if}{}{else if}{else}{end if}%
 
\setlist[itemize]{leftmargin=4mm}

\newcolumntype{L}[1]{>{\raggedright\let\newline\\\arraybackslash\hspace{0pt}}m{#1}}
\newcolumntype{C}[1]{>{\centering\let\newline\\\arraybackslash\hspace{0pt}}m{#1}}
\newcolumntype{R}[1]{>{\raggedleft\let\newline\\\arraybackslash\hspace{0pt}}m{#1}} 

\def\BibTeX{{\rm B\kern-.05em{\sc i\kern-.025em b}\kern-.08em
    T\kern-.1667em\lower.7ex\hbox{E}\kern-.125emX}}

\begin{document}
	
\title{Hybrid classical-quantum computing: are we forgetting the classical part in the binomial?\\
	{}
	\thanks{This work was supported %by the Basque Government through HAZITEK program (Q4\_Real project, ZE-2022/00033). This work was also supported 
    by the Spanish CDTI through Plan complementario comunicación cuántica (EXP. 2022/01341)(A/20220551).}
	\thanks{© 2023 IEEE. Personal use of this material is permitted. Permission from IEEE must be obtained for all other uses, in any current or future media, including reprinting/republishing this material for advertising or promotional purposes, creating new	collective works, for resale or redistribution to servers or lists, or reuse of any copyrighted component of this work in other works.}
}

\author{
	\IEEEauthorblockN{Esther Villar-Rodriguez\IEEEauthorrefmark{2}\IEEEauthorrefmark{1}, 
        Aitor Gomez-Tejedor\IEEEauthorrefmark{2}, and
	    Eneko Osaba\IEEEauthorrefmark{2}}
	\IEEEauthorblockA{\IEEEauthorrefmark{2}TECNALIA, Basque Research and Technology Alliance (BRTA), 48160 Derio, Bizkaia, Spain}
	\IEEEauthorblockA{\IEEEauthorrefmark{1}Corresponding author. Email: esther.villar@tecnalia.com}}
\maketitle

\IEEEpubidadjcol	

\begin{abstract}
The expectations arising from the latest achievements in the quantum computing field are causing that researchers coming from classical artificial intelligence to be fascinated by this new paradigm. In turn, quantum computing, on the road towards usability, needs classical procedures. Hybridization is, in these circumstances, an indispensable step but can also be seen as a promising new avenue to get the most from both computational worlds. Nonetheless, hybrid approaches have now and will have in the future many challenges to face, which, if ignored, will threaten the viability or attractiveness of quantum computing for real-world applications. To identify them and pose pertinent questions, a proper characterization of the hybrid quantum computing field, and especially hybrid solvers, is compulsory. With this motivation in mind, the main purpose of this work is to propose a preliminary taxonomy for classifying hybrid schemes, and bring to the fore some questions to stir up researchers minds about the real challenges regarding the application of quantum computing.

\end{abstract}

\section{Are we forgetting the classical part in hybrid classical-quantum computing?}\label{sec:abstract}

Quantum Computing (QC) has made great progress in recent years, mostly due to the rapid advancements in hardware and the democratization of its access. Consequently, different application fields have already benefited from the notable scientific progress made in QC to launch some proofs of concept. These attempts often rely on hybrid schemes, which represent the near-term future of this discipline.

In this regard, among other papers, two recent works have brought to the forefront the need of characterizing the hybrid computing world and order the community efforts in this field. These are their perspectives on the definition of hybrid \textbf{X} (we will later delve on the importance of clearing the $X$):
\begin{itemize}
    \item \textit{Hybrid \textbf{computing} as 'a global situation where we have a collection of computational tasks in which both the quantum and classical computers are used'} \cite{phillipson2023classification}. 
    \item \textit{A hybrid \textbf{algorithm} as 'an algorithm that requires nontrivial amounts of both quantum and classical computational resources to run, and which cannot be sensibly described, even abstractly, without reference to the classical computation'} \cite{callison2022hybrid}.
\end{itemize}

The reality is that the coexistence of classical and quantum methods will be perpetual in QC. In light of this, the heart of the matter is to reach consensus about what taxonomy, i.e arrangement of concepts, is more informative. 

In any case, it makes sense that the root of the classification tree be assigned to the hybrid \textit{what}. This means first \textit{distinguishing between a hybrid pipeline (HP) and a hybrid solver (HS)}. A \textbf{hybrid pipeline} is any workflow where classical and quantum processes take part (in line with \cite{phillipson2023classification}), whereas a \textbf{hybrid solver} is a particular case of a hybrid pipeline, implying that both the classical and quantum parts are involved in the core stages of a solver, i.e. searching and finding a complete, albeit binary, solution to the problem. 

Notwithstanding its extensive use, the definition of \textit{HS} has become too vague to account for nor the primary motivation behind the conjunction of these two paradigms in a pipeline or the individual contributions in the compound. This also hinders analysing prospective lines of study. On this basis, two categories are proposed with a key question to be answered: what is the role of the classic procedure within the HS?

\begin{itemize}
    \item \textbf{Supportive}. The classical procedures assist in the solving of a problem which would be, ideally, fully-intended for a quantum processor. In other words, traditional computation is used as a tool to make a QC-based solver feasible and/or more competent. Thus, it is a \textit{collaborative hybrid scheme} with no partnership in the key stages of the solver. 
    \item  \textbf{Intelligence sharing}. The HS contains a core intelligence engine composed of both classical and quantum artifacts. In other words, classical artificial intelligence mechanisms \textit{cooperate} with quantum-based modules synergistically.
\end{itemize}

\begin{figure}[t]
    \centering
    \includegraphics[width=0.95\linewidth]{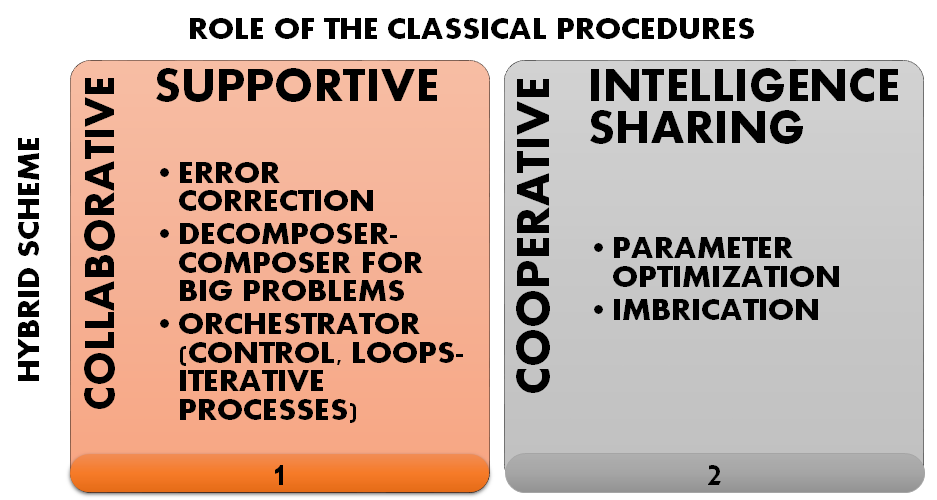}
    \caption{Classification of HS schemes.}
    \label{fig:tlp}
\end{figure}

With the aim of substantiating the basis of our proposal, this study is, from this point onwards, centered on one of the most exploited application of the classical methods in QC: \textit{decomposer-composer in collaborative HS} ((1) in Fig. \ref{fig:tlp}) and \textit{imbrication in cooperative HS} ((2) in Fig. \ref{fig:tlp}). For the sake of illustration, both categories are exemplified with D-Wave Systems approaches in the field of optimization. 

\textbf{Decomposer-composer in \textit{collaborative} HS}: The objective of these classical methods is two-fold: (1) break down the primal problem formulation into subproblems compatible with the quantum hardware limitations; and (2) build the complete solution from the partial results. In view of this, the main strategies in this area are contained in the following classification:
\begin{itemize}
    \item \textbf{Use Case-specific decomposers}: by leveraging partitioning strategies governed by heuristics specialized in the use case particularities. This category makes use of a \textit{vertical} partition, i.e. when the partition scheme is focused on dividing the set of variables through subproblem identification. This often involves building an initial hypothesis about the potential composition of the global optimum, i.e., detecting characteristics presumed in the optimal solution to break down the original problem into pieces whose partial solutions have a high chance of belonging to the optimal solution. In the Traveling Salesman Problem, using a distance-based clustering might be a promising decomposition strategy, based on the hypothesis that trips between neighboring points are prone to be part of the optimum solution \cite{osaba2021hybrid}.
         
    \item \textbf{General-purpose decomposers}: methods agnostic to the problem formulation rooted in heuristics of universal application. Habitually, these methods take advantage of a representation or model of the problem formulation that can be easily manipulated to be split into \textit{traversal partitions}. In the optimization field, the Energy-Impact and branch-and-bound algorithms incorporated into the D-Wave Hybrid framework are good exemplification of these strategies \cite{Hybrid}.
\end{itemize}

\textbf{Imbrication in \textit{cooperative} hybrid schemes}: with the aim of building a more sophisticated resolution scheme, classical and quantum procedures fuse their intelligence and work cooperatively to build a solution to a problem. This means that both participate in searching the solution of the problem. 
\begin{itemize}
    \item \textbf{Use Case-specific imbricated solvers}. This means resorting to \textit{horizontal partitions}, i.e. stratifying the problem formulation into several stages by virtue of problem-relaxation strategies. An example of this kind of decomposition can be found in \cite{feld2019hybrid} in the context of the Capacitated Vehicle Routing Problem.
    \item \textbf{General-purpose imbricated solvers}. Examples of these \textit{traversal partitioning} techniques are the well-known \textit{QBSolv}, in which the classical module builds a complete solution to a given problem while the quantum module takes different fractions of this outcome and performs different modifications on its structure \cite{boost2017partitioning}. The algorithm coined as \textit{Freeze and Anneal} \cite{bass2021optimizing} is another representative example that falls into this category, in which a classical genetic algorithm is in charge of building a preliminary partial solution to the problem to be finally completed by a quantum annealer solver.
\end{itemize}

Last but not least, every approach within these categories is inescapably coupled with a computational strategy. \textbf{Namely, vertical partitions could be processed in parallel threads since they are complete yet partial copies of the whole formulation, whereas sequential solving steps are the only alternative for the horizontal and traversal partitioning}. 

With the aforementioned taxonomy and examples in mind, we could enter into several discussions on the following topics and questions:
\begin{itemize}
    \item \textbf{HP}: when selecting an approach is a matter of time complexity, are we ensuring that the quantum advantage is not overridden by continuous calls between distant processors? 
    \item \textbf{HS}: will POST-NISQ era draw a different landscape regarding collaborative and cooperative HS? Will collaborative HS still remain in the picture?
    \item \textbf{HS-Supportive}: in consonance with the previous point, are we evaluating the classical procedures to avoid computationally intensive and slow classical methods which jeopardize the advantage?  
    \item \textbf{HS-decomposers}: is the goal of the collaborative solvers to attract use-case owners to play QC? Is there anyone researching on this?
    \item \textbf{HS-cooperative}: is the community evaluating the importance of the computational strategy into the benchmarking and tests?
    \item \textbf{HS-imbricated}: is the community discriminating and fairly measuring the contributions of the classical and quantum intelligence into the problem resolution?
    \item The community is constantly developing cooperative HSs, but \textbf{is the community working cooperatively in multidisciplinary research groups} (computer scientists, physicists, use-case owners...) to bring the best into HSs?
    \item \textbf{Are we forgetting the classical part in the binomial?}
\end{itemize}

\bibliographystyle{IEEEtran}

\end{document}